\documentclass[aps,prd,groupedaddress,showpacs,showkeys]{revtex4}
\usepackage{epsfig}
\pdfoutput=1
\newcommand{\be}[1]{\begin{equation} \label{(#1)}}
\newcommand{\eq}{\begin{eqnarray}}
\newcommand{\ee}{\end{equation}}
\newcommand{\en}{\end{eqnarray}}
\newcommand{\ba}[1]{\begin{eqnarray} \label{(#1)}}
\newcommand{\ea}{\end{eqnarray}}

\newcommand{\rf}[1]{(\ref{(#1)})}

\begin{document}

\title{  Heavy Sterile Neutrinos in Tau Decays and the MiniBooNE Anomaly}
\author{ Claudio Dib$^1$, Juan Carlos Helo$^1$, Martin Hirsch$^2$, Sergey Kovalenko$^1$, Ivan Schmidt$^1$  \vspace*{0.3\baselineskip}}
%

%
\affiliation{$^1$ Universidad T\'ecnica Federico Santa Mar\'\i a, \\
Centro-Cient\'\i fico-Tecnol\'{o}gico de Valpara\'\i so, \\
Casilla 110-V, Valpara\'\i so,  Chile \vspace*{1.2\baselineskip} \\
$^2$ 
AHEP Group, Instituto de F\'isica Corpuscular-C.S.I.C./Universitat de Val\'encia \\
Edificio de Institutos de Paterna, Apartado 22085, E-46071 Val\'encia, Spain \vspace*{0.3\baselineskip}\\}

\date{\today}

\begin{abstract}

Current results of the MiniBooNE experiment show excess events that indicate neutrino oscillations, but only if one goes beyond the standard 3 family scenario. Recently a different explanation of the events has been given, not in terms of oscillations but by the production and decay of a massive sterile neutrino with large transition magnetic moment. We study the effect of such a sterile neutrino in the rare decays $\tau^- \rightarrow \mu^-  \mu^+ \pi^- \nu$ and $\tau^{-}\rightarrow \mu^{-}  \mu^{+} e^{-} \nu  \nu$. We find that searches for these decays featuring displaced vertices between the $\mu^-$ and the other charged particles, constitute good tests for the existence of the sterile neutrino proposed to explain the MiniBooNE anomaly. These searches could be done with already existing experimental data.

\end{abstract}



\maketitle

\newpage

In 1996 the LSND experiment presented evidence of $\bar \nu_\mu \rightarrow \bar \nu_e$ which was not consistent with the global neutrino data \cite{LSND}. This experiment used $\bar \nu_\mu$ produced in the beam stop of a proton accelerator. The $\bar\nu_\mu$ energy distribution peaked near 55 MeV, with a mean energy near 100 MeV. The search for $\bar \nu_\mu \rightarrow \bar \nu_e$ was based on the appearance of $\bar \nu_e$ in the neutrino beam, detected through the reaction $\bar \nu_e p^+ \rightarrow e^+ n$, resulting in a relativistic $e^+$ \cite{LSND}. With the aim  to confirm this anomaly, the MiniBooNE experiment was designed to search for $ \nu_\mu \rightarrow  \nu_e$,  at higher energies  than the LSND experiment (namely above 475 MeV, peaked near 600 MeV and average near 800 MeV), but at similar $L/E$, $L$ being the distance travelled by the neutrinos and $E$ their energy. In this search the MiniBooNE experiment did not find positive signals \cite{MiniBooNE_NF}. However, looking at  lower energies (below 475 MeV) the MiniBooNE collaboration has observed an excess of electron-like events in the energy distribution of charge-current quasi-elastic electron neutrino events \cite{MiniBooNE}.  Recently, the MiniBooNE collaboration, in searches $\bar \nu_\mu \rightarrow \bar \nu_e$ oscillations,  have also found that the antineutrino data have an anomalous low energy excess similar to that of the neutrino data \cite{MiniBooNEUpdate}.

Clarification of the MiniBooNE and LSND anomaly is a very important task. Over the last decade, many explanations
have been proposed, including new oscillation physics \cite{Valle}  and production of photons in neutrino scattering
events \cite{Gninenko, Gninenko40}. Since the MiniBooNE and LSND  detectors cannot differentiate between Cerenkov rings produced by electrons (positrons) or converted photons, the excess observed could come from photons instead of electrons (positrons) as assumed in the neutrino oscillation paradigm. Consequently,  one of the main  problems is that  we do not know whether the MiniBooNE and LSND excess events are both signals of neutrino oscillation, or if any of them has a different origin.

The MicroBooNE  experiment is being planned to start taking data next year in order to check the MiniBooNE anomaly \cite{MicroBooNE}. It can
confirm the MiniBooNE excess as well discriminate between production of electrons or photons,
hence clarify the nature of the excess \cite{MicroBooNE}.
If the signals turn out to be from photons,  they will not be due to neutrino oscillation but may come from the dominant radiative decay of a
sterile neutrino $N$ with mass $m_{N}$, mixing strength $U_{\mu N}$  and lifetime  $\tau_{N}$ in the range: \cite{Gninenko,  GninenkoP}
\ba{Range}
400\textrm{ MeV} \lesssim m_N \lesssim 600\textrm{ MeV}  \ \  \  , \ \ \
1 \times 10^{-3} \lesssim |U_{\mu N}|^2 \lesssim 4 \times 10^{-3}   \ \  \  , \ \ \
\tau_N \lesssim 1 \times 10^{-9} \textrm{ s} .
\ea
This sterile neutrino is produced by $Z^0$ exchange of
the incoming $\nu_\mu$ or $\bar\nu_{\mu}$ with a nucleus in the detector \cite{Gninenko}.
In turn, the subsequent radiative decay of $N$ requires a transition magnetic moment within the following range, in order to explain the MiniBooNE signal:
  \ba{mu}
  \mu_{tr} \simeq (1-6)\times  10^{-9} \mu_B,
  \ea
 where $\mu_B$ is the  Bohr  magneton. These values imply that $N$ decays dominantly in the mode $N\rightarrow \nu \gamma$,  and, therefore, the total width $\Gamma_N$ can be approximated by \cite{Mohapatra}
\ba{DecayMode}
\Gamma_N \sim \Gamma(N\rightarrow \nu \gamma) = \frac{\alpha}{8} \left(\frac{\mu_{tr}}{\mu_B}\right)^2 \left(\frac{m_N}{m_e}\right)^2 m_N,
\ea
thus constraining the sterile neutrino lifetime shown in Eq.\ \rf{Range} to be in the range 
\ba{Range2}
2 \times 10^{-11} \textrm{ s} \lesssim \tau_N \lesssim 1 \times 10^{-9} \textrm{ s}.
\ea
Of course, this explanation suggests that the MiniBooNE anomaly comes from a different physics than the LSND \cite{LSND} and other neutrino anomalies \cite{Guinti2}. Motivations along this line have been proposed in \cite{Guinti}, where it is  found that the appearance and disappearance data are marginally compatible in a (3+1) neutrino mixing model, if  we disregard the MiniBooNE data of the
low-energy anomaly. Then, according to \cite{Guinti},  the MiniBooNE excess  seems to  have an explanation other than  neutrino oscillations.

%

There have also been attempts to explain  the MiniBooNE and LSND anomalies simultaneously using a similar scenario with   a radiative decaying sterile neutrino $N\rightarrow \nu \gamma$,  this time using sterile neutrino masses around $60$ MeV \cite{Gninenko40} (see also \cite{RefG}). However, such scenario has been recently  ruled out using direct searches of radiative $K$ meson decays at ISTRA+ Setup \cite{ISTRA}.

Concerning experimental  tests of a sterile neutrino in the range given by the Eqs.\  \rf{Range}, \rf{mu} and \rf{Range2}, direct searches using $D_s$ decays have already been proposed \cite{GninenkoDs}. Morover, the MicroBooNE experiment could easily probe this model if they find that the MiniBooNE anomaly is due to photons and not electrons. The main purpose of this paper is to propose searches for a heavy neutrino in $\tau$ decays, with properties described in \cite{Gninenko} and parameters in the range given by Eqs.\  \rf{Range}, \rf{mu} and \rf{Range2}.

\begin{figure}[htbp]
\centering
\includegraphics[width=0.8\textwidth,bb=70 650 500 775]{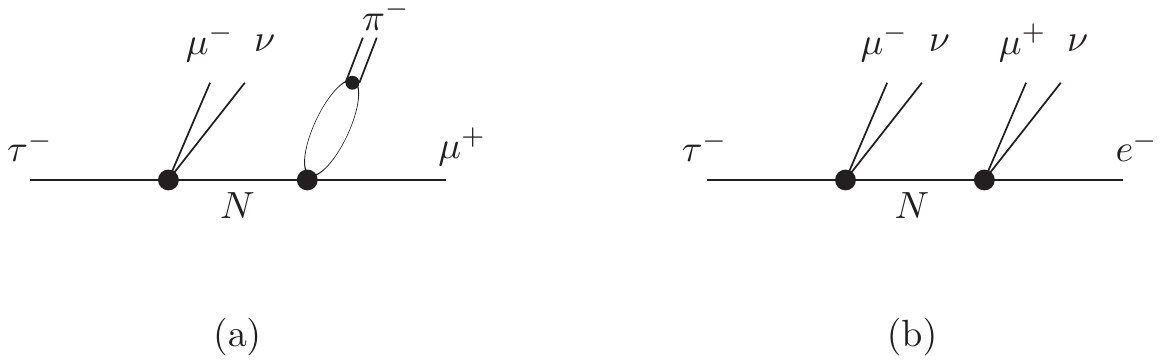}
\caption{Structure of the on-shell sterile neutrino $N$  contribution  to the  $\tau$  decays \rf{DECAYS}.
}
\label{fig-1}
\end{figure}

If a sterile neutrino $N$ with parameters in the range   \rf{Range}, \rf{mu} and \rf{Range2} exists, it should contribute as an on-shell intermediate particle in the following  $\tau$ decays:
 \ba{DECAYS}
\tau^- \rightarrow \mu^-  \mu^+ \pi^- \nu  \ \ \  \textrm{and} \ \  \ \tau^{-}\rightarrow \mu^{-}  \mu^{+} e^{-} \nu  \nu  .
\ea
  This means that an intermediate sterile neutrino is produced at the corresponding vertex on the left of the diagrams in Fig.~\ref{fig-1}, propagates as a free unstable particle, and then decays at the corresponding vertex on the right.  These decays could be searched in $\tau$ decay events with two vertices leading to very clean signals:  a primary vertex from $\tau^- \rightarrow \mu^- \nu N$
and a secondary displaced vertex  from $N \rightarrow \mu^+ \pi^- ( \mu^+ e^- \nu)$. In the case of $\tau^- \rightarrow \mu^-  \mu^+ \pi^- \nu $, the experimental signature $\mu^+ \pi^-$ coming from the  displaced vertex would be two charged tracks and, since there is no neutrino in the final state, it could be  possible to reconstruct the mass of the sterile neutrino \rf{Range}. This would show as a peak in invariant mass squared of the pair in the range $0.16-0.36$ GeV$^2$, thus providing an excellent cross-check of the model.

The  $\tau$ decay rates in question, dominated by an on-shell sterile neutrino in the intermediate state (see Fig.~\ref{fig-1}) are:
\begin{eqnarray}\label{DECRATE}
 \Gamma^{ }(\tau^{-}\rightarrow \mu^{-}  \mu^{+} e^{+} \nu  \nu)&\approx&
\delta_{N} \cdot \Gamma(\tau^{-}\rightarrow \mu^{-}  \bar{\nu}_\mu N)\frac{\Gamma( N \rightarrow  \mu^+ e^-  \bar{\nu}_e)}{\Gamma_N} +
  \Gamma(\tau^{-}\rightarrow \mu^{-}  \nu_\tau N)\frac{\Gamma( N \rightarrow \mu^+ e^-  \bar{\nu}_e)}{\Gamma_N}
    \\  \Gamma^{ }(\tau^{-}\rightarrow \mu^{-}  \mu^{+}  \pi^-  \nu)&\approx&  \delta_{N} \cdot \Gamma(\tau^{-}\rightarrow \mu^{-}  \bar{\nu}_\mu N)\frac{\Gamma( N \rightarrow  \mu^+ \pi^-  )}{\Gamma_N} + \Gamma(\tau^{-}\rightarrow \mu^{-}  \nu_\tau N)\frac{\Gamma( N \rightarrow \mu^+ \pi^- ) }{\Gamma_N}.
 \end{eqnarray}
   Here  $\delta_{N} = 0,1$ for the Dirac and Majorana cases of the sterile neutrino $N$, respectively.  For simplicity we have used $U_{e N} = 0$ which is reasonable considering that $U_{e N}$ is strongly constrained in the  sterile neutrino mass range of Eq.\ \rf{Range}  \cite{Atre:2009rg}. The  $\tau$ and $N$ partial decay rates are \cite{K-decPaper, Atre:2009rg}
    \begin{eqnarray}\label{DECAYRATETAU}
\Gamma(\tau^- \rightarrow \mu^{-}\nu_{\tau} N )&=& |U_{\mu N}|^2
\frac{G_F^2}{192\pi^3} m_\tau^5 I(z_{N},z_{\nu}, z_{\mu}), \ \ \ \ \ \ \ \ \
\Gamma(\tau^- \rightarrow \mu^{-}\bar \nu_{\mu} N )= |U_{\tau N}|^2
\frac{G_F^2}{192\pi^3} m_\tau^5 I(z_{N},z_{\nu}, z_{\mu}),
\\
\label{DECAYRATETAU2}
\Gamma(N\rightarrow e^{-}\mu^{+}\bar \nu_{e} )&=& |U_{\mu  N}|^2
\frac{G_F^2}{192\pi^3} m_N^5 I(y_{e },y_{\nu}, y_{\mu}) , \ \ \ \ \ \ \ \ \
%
%
\label{lP}
\Gamma(N\rightarrow \mu^+ \pi^{-}) = |U_{\mu N}|^2
\frac{G_F^2}{16 \pi}m_N^3 f_{\pi }^2 |V_{ud }|^{2} F_\pi (y_{\mu },y_{\pi }).
%
\end{eqnarray}
Here we have defined $z_{i} = m_{i}/m_{\tau}$, $y_{i} = m_{i}/m_{N}$, for $m_{i} = m_{N}, m_\nu, m_\mu, m_e, m_\pi$;    $f_{\pi}=130$ MeV.
The kinematical functions in Eqs. (\ref{DECAYRATETAU})-(\ref{lP}) are
\ba{kin-fun-1}
 &&I(x,y,z)= 12 \int\limits_{(x+y)^{2}}^{(1-z)^{2}} \frac{ds}{s}
(s-x^2-y^{2})(1+z^2-s) \lambda^{1/2}(s, x^{2}, y^2) \lambda^{1/2}(1, s, z^2),
\\
&&F_P(x,y)= \lambda^{1/2}(1,x^2,y^2) [(1+x^2)(1+x^2-y^2) - 4 x^2].
%
%
\ea
The corresponding $\tau$ branching ratios, for the parameters in Eqs.\ \rf{Range}, \rf{mu},  are in the range:
 \ba{BrN}
Br(\tau^- \rightarrow \mu^-  \mu^+ \pi^- \nu) = 8.2 \times 10^{-5} - 2.1 \times 10^{-8}
 \ \ \ \ \ \ ; \ \ \ \ \ \ Br( \tau^{-}\rightarrow \mu^{-}  \mu^{+} e^{-} \nu  \nu) = 1.3 \times 10^{-5} - 2.0 \times 10^{-9}.
  \ea
In these numerically evaluations we have used $|U_{\tau N}|^2 < 10^{-2}$,
 which is consistent with the best current limits in the sterile neutrino mass range \rf{Range}  \cite{Atre:2009rg,UTAU}.

Another issue to take into account in the $\tau$ decays  of Eq.\ \rf{DECAYS} concerns the probability  $P_{N}$ for the neutrino $N$ to decay inside the detector.   Roughly,  for a  detector of length $L_{D}$, the probability  $P_{N}$ takes the form $ P_{N}\approx1-e^{-{L_{D}}/{L}}$, where  $L=  \ \gamma c \ \tau_N  $  is the  decay length of the sterile neutrino. Assuming $\gamma =1$, the  decay length  $L $, within which $\sim 63\%$ of the sterile neutrinos decay,  is
\ba{L}
L =  0.6 - 30 \textrm{ cm} .
\ea
  The most sensitive detectors  to the $\tau$ decays in Eq.\ \rf{DECAYS} are  SuperB \cite{SUPERB}, Belle \cite{BELLE},   BaBar \cite{BABAR}, CLEO-c \cite{CLEO-C} and BES-III \cite{BESS-III}. In particular BaBar and Belle have  $4.9 \times 10^{8}$ and $7.2 \times 10^{8}$ $\tau^+ \tau^-$ pairs
of events, respectively \cite{BFactories}, which for clean signals correspond to a sensitivity for the $\tau$ branching ratios of order (few) $O(10^{-9})$.

Then, considering that the decay length $L$ in Eq.\ \rf{L} is smaller than the detector sizes, and the branching ratios in Eq.\ \rf{BrN} are within the reach of Babar and Belle, we conclude that displaced vertex searches of $\tau$ decays in Eq.\ \rf{DECAYS} should be able to test the sterile neutrino scenario corresponding to the parameters given in Eqs.\  \rf{Range}, \rf{mu} and \rf{Range2}, proposed as a non-oscillation explanation of the MiniBooNe anomaly \cite{Lusiani}.

In summary, we suggested that searches for the rare $\tau$ decays $\tau^- \rightarrow \mu^-  \mu^+ \pi^- \nu$ and
$\tau^{-}\rightarrow \mu^{-}  \mu^{+} e^{-} \nu  \nu$ exhibiting displaced vertices (i.e. a primary vertex where $\mu^-$ is produced and a secondary vertex where the pair $\mu^+ \pi^-$ or $\mu^+ e^-$ is produced, respectively), should constitute tests for the existence of a massive sterile neutrino
with parameters in the range shown in Eqs.\  \rf{Range}, \rf{mu} and \rf{Range2}, required to explain the MiniBooNE anomaly without neutrino oscillations. These searches could be done with already existing experimental data.



\begin{center}
{\bf Acknowledgements}
\end{center}

We are grateful to  Ignacio Aracena, Hayk Hakobyan and Will Brooks for useful discussions. We also thank Sergey 
Gninenko and Alberto Lusiani for useful comments. J.C.H. thanks the IFIC for hospitality during his stay.
This work was supported by FONDECYT (Chile) under projects 1100582, 1100287; 
Centro-Cient\'\i fico-Tecnol\'{o}gico de Valpara\'\i so PBCT ACT-028, 
by Research Ring ACT119, CONICYT (Chile) and CONICYT/CSIC 2009-136. M.H. acknowledges support from the Spanish MICINN grants FPA2008-00319/FPA, FPA2011-22975, MULTIDARK CSD2009-00064 and
2009CL0036 and by CV grant Prometeo/2009/091 and the
EU~Network grant UNILHC PITN-GA-2009-237920.



\begin{thebibliography}{99}
\bibitem{GninenkoP} The model is in agreement with the latest MiniBooNE results; Sergei Gninenko, private       
                                            communication.
\end{thebibliography}
\end{document}